# Analysis of persistence-based solar irradiance forecasting benchmarks

*Rodrigo Alonso-Suárez, Daniel Aicardi, Franco Marchesoni-Acland*

**Summary**

In this work we analyse a set of benchmark methods for solar irradiance forecasting based on the clear-sky index, namely, persistence, climatology, smart-persistence and convex combination (CC) of persistence and climatology. To assess the boundaries of the benchmarks, we include in the analysis a simple least squares regression technique that uses the last five past samples. For the methods that need data adjustment, we also analyze their variation with different training sets. This work is done for intra-day forecast of global solar horizontal irradiance with 10-minutes and hourly forecast horizons. Our results, obtained in the south-east region of South America known as Pampa Húmeda, confirm the benchmark recommendation, showing that the CC method is exigent for RMSD as it achieves almost the same performance of the more complex linear combination of past measurements. We also observe that the benchmarks have different behavior depending on the metric, as the most exigent benchmark for one metric is not necessarily the most exigent for another.

## 1. Introduction

Solar irradiance forecasting is important to accelerate the inclusion of solar energy sources in the world's energy supply. This problem is formulated in diverse ways, including different time horizons (intra-hour, intra-day, days-ahead, seasonal, etc.), different worldwide locations and climates, different methods (Diagne et al., 2013; Antonanzas et al., 2016), different models' inputs (ground measurements and images, satellite data, etc.) and different targets (global or direct irradiance, photovoltaic power output, etc.). There are mainly four approaches to tackle this problem and their hybrid combinations, which are best suited for different settings: i) Numerical Weather Predictions (NWP), ii) satellite-based forecasts, iii) time series forecasting methods, and iv) computer vision methods or algorithms using data from ground cameras. Forecasts are usually evaluated based on some well-known metrics (MBD, MAD, RMSD), but a benchmark method is required to calculate more meaningful metrics as the Forecasting Skill (FS) score, which represents the improvement given by any method when compared with the benchmark. The need for a benchmark arises from the meteorological diversity, which influentiates the variability of the solar resource and, consequently, the performance of any method. A recent article authored by many renowned researchers in the field establishes criteria for this benchmark and performance metrics (Yang et al., 2020), strongly recommending the use of the RMSD based FS score and the convex combination of persistence and climatology as benchmark (Yang et al. 2019), among other relevant analysis. Here, we aim to contribute to this issue by providing a regional comparison of different benchmark strategies based on the clear-sky index, including the recommended convex combination, for a region with intermediate solar variability and in two intra-day time scales (hourly and 10-minutes). The comparison is different depending on the time basis of the forecast, but we verify that for both scales the convex combination is the best choice, being simple and exigent. This benchmark obtains almost the same performance (in RMSD) that a linear combination of recent past samples. We also notice that benchmarks that are not exigent in terms of RMSD can be exigent in terms of MAD. Finally, we also analyze the effect of using different data sets for adjusting required parameters.

## 2. Data

For this work we use solar global horizontal irradiance (GHI) at 4 sites in the Pampa Húmeda region of the south-east of South America, a region classified as Cfa in the updated Köppen-Geiger climate classification (Peel et al., 2007) with intermediate short-term solar irradiance variability (Alonso-Suárez et al., 2020). The data was registered with calibrated spectrally-flat class A Kipp & Zonen pyranometers (ISO 9060:2018) and was quality checked at a 1-minute time-rate by using the BSRN filters with local coefficients (McArthur, 2005). We use the data for two complete years (2019-2020). The evaluation of the methods is done by using the 2020 measurements. As some methods (all except regular/smart persistence) require historical data to tune parameters, we evaluate them in two settings: (i) the pretrained setting, in which the year 2019 is used for training, and (ii) the data-snooping setting, in which the year 2020 is used for training. The performance of the operational setting, in which the training is done online with the historical data available at the forecast issue time, will lie between (i) and (ii).

## 3. Methods and results

We implement four benchmark procedures for GHI based on the clear-sky index ($k_c$), and compare them with a simple forecasting method based on a static linear combination of recent past samples (LS). The considered benchmarks are: (a) the persistence ($k_c(t+h) = k_c(t)$), with $h$ being forecast horizon, (b) the smart persistence with varying $n$ ($k_c(t+h) = \sum_i k_c(t-i) / n$ for $i$ in $\{0, n-1\}$), (c) the climatological value, and (d) the convex combination (CC) of (a) and (c) as proposed by Yang et al. (2019). For the LS method we used the current time and 4 past samples (5 in total, calling it LS-5), as this has been the optimal number of lags in previous studies (Lauret et al., 2005; Marchesoni & Alonso-Suárez, 2020). The clear-sky index is calculated with the McClear model (Lefèvre et al., 2013), retrieved from the Copernicus Atmosphere Monitoring Service.

The comparison results are outlined in Fig. 1, taking the data-snooping case as an example for this abstract. The relative RMSD is shown in the panels at the left and the relative MAD in the panels at the right. The top panels correspond to the hourly time basis and the bottom ones to the 10-minutes time basis. The smart persistence is shown in yellow and the curves are located higher with increasing $n$ for the first time horizon. The behavior of the procedures is different depending on the metric and the time scale.

For hourly scale (Fig. 1, top), the CC benchmark is by far the best alternative in terms of RMSD, achieving almost the same levels as the LS method and outperforming the rest of the benchmarks for all time horizons. The smart persistence at hourly scale does not have any advantage in RMSD, as for the shorter time horizons does not outperform the regular persistence and for the larger does not outperform the climatology. The MAD hourly plot shows that the simple $k_c$ persistence benchmark is competitive up to 3-hours ahead for this metric, when is not an exigent option for RMSD. The climatology is not a good option in any scenario, being significantly worse for the MAD metric than for the RMSD.

The analysis for the 10-minutes time frame is rather different (Fig. 1, bottom). The CC method is still the best benchmark, but the smart persistence tights, having a similar RMSD approximately for the first hour of forecast. The LS-5 algorithms slightly outperform CC, but only in the shorter time horizons, showing that the benchmark is still exigent under this time basis. The smart persistence is indeed an option for the MAD metric. The lowest MAD curve is obtained by taking the smart persistence with the optimal $n$ averaging value for each time horizon, which for instance is $n=1$ for the first time step (and increases for the following). For the MAD metric the simple kc persistence is, again, a competitive benchmark, but only up to ~3 hours ahead. In the same way as for the hourly case, climatology is not a good option for any metric.

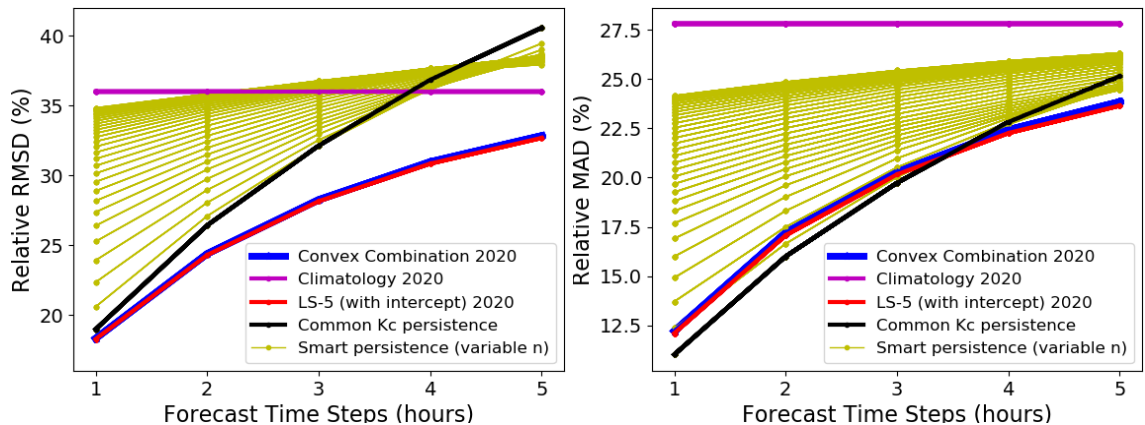

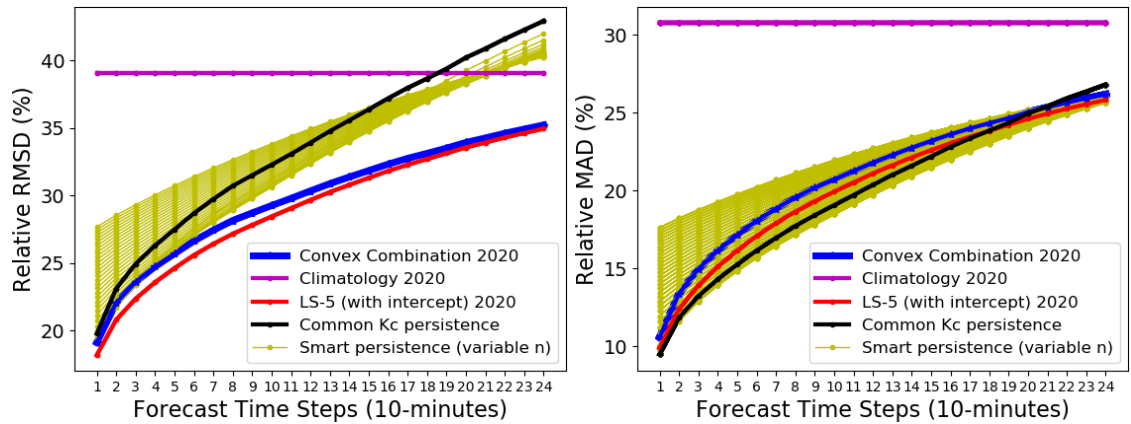

Fig. 1: Performance comparison as a function of the forecast horizon. Top: hourly time basis. Bottom: 10-minutes time basis.

## 4. Conclusions

This work seeks to contribute with the understanding of solar forecasting benchmarks, evaluating several naive methods for a region with intermediate solar short-term variability. The assessment is done for the three main metrics (MBD, MAD and RMSD) and for two different intra-day time scales. An important requirement for comparability between forecasting proposals is the adoption of a benchmark method to use. There is then a compromise between adopting a new, more performant method, and sticking to one that enjoys widespread use. In this sense, this work verifies for the Pampa Húmeda region that the CC method, being simple and exigent, is the best option for such widespread baseline, as recommended. Also, we highlight some differences of the commonly used benchmarks when considered for other metrics rather than the RMSD. The complete article will also include the discussions regarding the MBD metric and the training set. This last analysis is, a priori, relevant for the CC and climatology benchmarks (and the LS-5 technique). We found slight variations in these benchmarks by using different training sets, which is a desired property.